\documentclass[12pt]{article}
\usepackage{epsfig}
\usepackage{latexsym}
\begin{document}
\title{Quasinormal modes for Weyl neutrino field in R-N black holes}
\author{\begin{tabular}{c}\bigskip Hongbao Zhang$^{1,2}$\footnote{Email: hongbaozhang@pku.edu.cn}, Zhoujian Cao$^{3}$, Xuefei Gong$^3$, Wei Zhou$^{3}$\\  \smallskip$^1$School of Physics, Peking University, Beijing, 100871, PRC
\\ \smallskip$^2$Institute of Applied Mathematics, Academia Sinica,
Beijing, 100080, PRC \\$^3$Department of Physics, Beijing Normal
University, Beijing, 100875, PRC\end{tabular}}
\maketitle
\begin{abstract}
We employ WKB approximation up to the third order to determine the
low-lying quasinormal modes for Weyl neutrino field in R-N black
holes, which are the most relevant to the evolution of the field
around a black hole in the intermediate stage. It is showed that
the quasinormal mode frequencies for Weyl neutrino field in R-N
black holes are different from those in Schwartzchild black holes
 owning to the charge-induced additional gravitation, and the variations of the quasinormal mode frequencies for Weyl
neutrino field are similar to those for integral spin fields in
R-N black holes.
\end{abstract}
\section{Introduction}
It is well known that quasinormal modes(QNM), which dominate in
the intermediate stage of field evolution around a black hole,
play a two-fold role in classical black hole physics: One is that
the existence of QNM ensures the stability of black holes against
small perturbations, because the small perturbations will damp
following QNM. The other is that a black hole can be identified
with its QNM; for they are "footprints" of black holes, since
their frequencies only depend on the fundamental parameters of
black holes such as mass, charge and angular
momentum\cite{Chandra,KS,Nollert}. The latter is quite significant
from the viewpoint of astrophysical observation because we could
determine the fundamental parameters of black holes by detecting
their QNM. Recently QNM have also shed light on the quantized area
of black holes in quantum gravity such as loop quantum gravity
\cite{Hod,Dreyer,Corichi,LZ,P,Swain,O}, which leads to a renewed
interest in QNM of black holes.

There has been much work done on calculation of QNM for different
fields around various black
holes\cite{KS1,VL,Konoplya1,BK,Cho,Konoplya2,Suneeta,Molina,Konoplya3,CA,M,Pa,CP};
however, as far as Weyl neutrino field in charged black holes,
namely R-N black holes, is concerned, no work has been done.
Nevertheless, with the development of the technology in neutrino
detection, the investigation of this case will aquire a practical
application in the astrophysical observation: Neutrino fluxes from
a star take a more and more important position in our
investigating the interior information of the star because
neutrino only takes part in the weak interaction except the
universal gravitation\cite{W,K}; especially when a sufficiently
massive charged star collapse to a black hole\cite{Pons}, the
neutrino fluxes can provide us with the information about the
fundamental parameters of the formed black hole due to their QNM
behaviors. In addition, this work is of interest itself: In spite
of the absence of charge-charge electromagnetic interaction
between neutrino and a R-N black hole, the charge of the R-N black
hole will influence the evolution of neutrino by way of the
charge-induced gravitation, which will cause the QNM frequencies
of Weyl neutrino field to deviate from those in Schwarzschild
black holes\cite{Cho}. The present paper serves to work on QNM for
Weyl neutrino field in R-N black holes.

In next section, we consider Weyl neutrino field equation in R-N
black holes in terms of N-P formalism and its reduction into a
pair of one-dimensional equations with supersymmetric partnership.
In Section 3, we calculate the low-lying QNM frequencies for Weyl
neutrino field using WKB approximation. The conclusion and
discussion are presented in Section 4.
\section{Weyl neutrino field equation in R-N black holes}
Weyl neutrino field equation in curved spacetime\cite{Wald}
\begin{equation}
\bigtriangledown_{A'A}\psi^A=0
\end{equation}
can be written as\cite{Chandra}
\begin{equation}
(D+\rho-\varepsilon)\psi^{1}+(\delta+\alpha-\pi)\psi^{2}=0\label{1}
\end{equation}
\begin{equation}
(\bar{\delta}+\tau-\beta)\psi^{1}+(\Delta+\gamma-\mu)\psi^{2}=0\label{2}
\end{equation}
With the usual form of the R-N black holes with parameters $M$ and
$Q$
\begin{equation}
ds^{2}=\frac{\Delta}{r^{2}}dt^{2}-\frac{r^{2}}{\Delta}dr^{2}-r^{2}(d\theta^{2}+\sin^{2}\theta
d\varphi^{2})
\end{equation}
where $\Delta=r^{2}-2Mr+Q^{2}$, if construct the null tetrad-frame
in N-P formalism as follows\footnote{In our notation, the
directional derivatives are the complex conjugate of those in
\cite{Chandra}; and the spin coefficients are the minus of those
in \cite{Chandra}.}
\begin{equation}
l^\mu=(l^t,l^r,l^\theta,l^\varphi)=\frac{1}{\Delta}(r^2,\Delta,0,0)\label{Derivative}
\end{equation}
\begin{equation}
n^\mu=(n^t,n^r,n^\theta,n^\varphi)=\frac{1}{2r^2}(r^2,-\Delta,0,0)
\end{equation}
\begin{equation}
m^\mu=(m^t,m^r,m^\theta,m^\varphi)=\frac{1}{\sqrt{2}r}(0,0,1,-i\csc\theta)
\end{equation}
then non-zero spin coefficients satisfy
\begin{equation}
\rho=\frac{1}{r},\alpha=-\beta=\frac{\cot\theta}{2\sqrt{2}r},\mu=\frac{\Delta}{2r^3},\gamma-\mu=\frac{M-r}{2r^2}\label{Spin}
\end{equation}
Therefore, if let
$\psi^{1}=\frac{1}{r}R_{-}(r)S_{-}(\theta)e^{-i\omega
t}e^{im\varphi},\psi^{2}=\frac{1}{\sqrt{\Delta}}R_{+}(r)S_{+}(\theta)e^{-i\omega
t}e^{im\varphi}$, substitute $(\ref{Derivative})-(\ref{Spin})$ to
(\ref{1}), (\ref{2}), and use
$\frac{dr_*}{dr}=\frac{r^2}{\Delta}$\footnote{$r_*$ is the
tortoise coordinate, defined in \cite{Chandra}.}, we have
\begin{equation}
(-i\omega
+\frac{d}{dr_*})R_-(r)=\lambda_1\frac{\sqrt{\Delta}}{r^2}R_+(r),\frac{1}{\sqrt{2}}(m\csc\theta+\frac{1}{2}\cot\theta+\frac{d}{d\theta})S_+(\theta)=-\lambda_1S_-(\theta)\label{Separate1}
\end{equation}
\begin{equation}
(i\omega
+\frac{d}{dr_*})R_+(r)=\lambda_2\frac{\sqrt{\Delta}}{r^2}R_-(r),\sqrt{2}(m\csc\theta-\frac{1}{2}\cot\theta-\frac{d}{d\theta})S_-(\theta)=-\lambda_2S_+(\theta)\label{Separate2}
\end{equation}
where $\lambda_1,\lambda_2$ are separate constants. Without loss
of generalization and for the sake of convenience, we let
$\lambda_1=\lambda_2=\lambda$. The explicit expression of
$\lambda$ can be obtained by combining the angular equations in
(\ref{Separate1}) and (\ref{Separate2})
\begin{equation}
[\frac{1}{\sin\theta}\frac{d}{d\theta}(\sin\theta\frac{d}{d\theta})-(\frac{m^2+\frac{1}{4}+m\cos\theta}{\sin^2\theta})]S_+(\theta)=-(\lambda^2-\frac{1}{4})S_+(\theta)\label{Spin1}
\end{equation}
\begin{equation}
[\frac{1}{\sin\theta}\frac{d}{d\theta}(\sin\theta\frac{d}{d\theta})-(\frac{m^2+\frac{1}{4}-m\cos\theta}{\sin^2\theta})]S_-(\theta)=-(\lambda^2-\frac{1}{4})S_-(\theta)\label{Spin2}
\end{equation}
The equations (\ref{Spin1}) and (\ref{Spin2}) have as their
solutions the spin-weighted spherical harmonics\cite{Goldberg}
\begin{equation}
S_{\pm}(\theta)=_{\pm\frac{1}{2}}Y^l_m(\theta),\lambda^2=(l+\frac{1}{2})^2
\end{equation}
where $l=\frac{2k-1}{2}$ with $k$ positive integers, and $-l\leq
m\leq l$. Without loss of generalization, we take $\lambda=k$ in
the following discussion.

Now setting $Z_{\pm}(r)=R_+(r)\pm R_-(r)$, and combining the
radial equations in (\ref{Separate1}) and (\ref{Separate2}), we
have
\begin{equation}
(\frac{d}{dr_*}-\lambda\frac{\sqrt{\Delta}}{r^2})Z_+(r)=-i\omega
Z_-(r)\label{supersymmetry1}
\end{equation}
\begin{equation}
(\frac{d}{dr_*}+\lambda\frac{\sqrt{\Delta}}{r^2})Z_-(r)=-i\omega
Z_+(r)\label{supersymmetry2}
\end{equation}
Finally we readily obtain a pair of one-dimensional equations
\begin{equation}
(\frac{d^2}{dr_*^2}+\omega^2)Z_{\pm}=V_\pm Z_\pm \label{WKB}
\end{equation}
where
\begin{equation}
V_\pm=\lambda^2\frac{\Delta}{r^4}\pm\lambda\frac{d}{dr_*}(\frac{\sqrt{\Delta}}{r^2})
\end{equation}
When $Q=0$, it is easy to show that the above effective potentials
$V_\pm$ reduce to those for Schwarzchild black holes in
\cite{Cho}.
\section{QNM for Weyl neutrino field}
In this section, we shall only concentrate on calculating the
low-lying QNM for Weyl neutrino field using WKB approximation,
since the low-lying modes dominate in the evolution of Weyl
neutrino field in the intermediate stage.

Note that the equations with respect to $Z_+$ and $Z_-$ in
(\ref{WKB}) possess the same spectra of QNM, because $Z_+$ and
$Z_-$ are related as supersymmetric partnership through
(\ref{supersymmetry1}) and (\ref{supersymmetry2}); therefore to
calculate the spectra of QNM, we need only focus on the equation
with $Z_+$ in (\ref{WKB})
\begin{equation}
(\frac{d^2}{dr_*^2}+\omega^2)Z=VZ \label{WKB1}
\end{equation}
where
\begin{equation}
V=\lambda^2\frac{\Delta}{r^4}+\lambda\frac{d}{dr_*}(\frac{\sqrt{\Delta}}{r^2})\label{V}
\end{equation}
here we have written $V,Z$ as $V_+, Z_+$. Furthermore, according
to (\ref{supersymmetry1}) and (\ref{supersymmetry2}), we find that
the QNM frequencies with ${\rm Re}(\omega)<0$ are related to those
with ${\rm Re}(\omega)>0$ by a reflection via the imaginary axis
in the complex plane\footnote{Here it is worth pointing that the
QNM spectra for anti-neutrino field in any black hole is the same
as that for neutrino field because their solutions are related by
the complex conjugation\cite{Wald}.}. Thus we need only calculate
the QNM frequencies with ${\rm Re}(\omega)>0$.

\begin{figure}
\begin{center}
\scalebox{0.8}[0.8]{\includegraphics{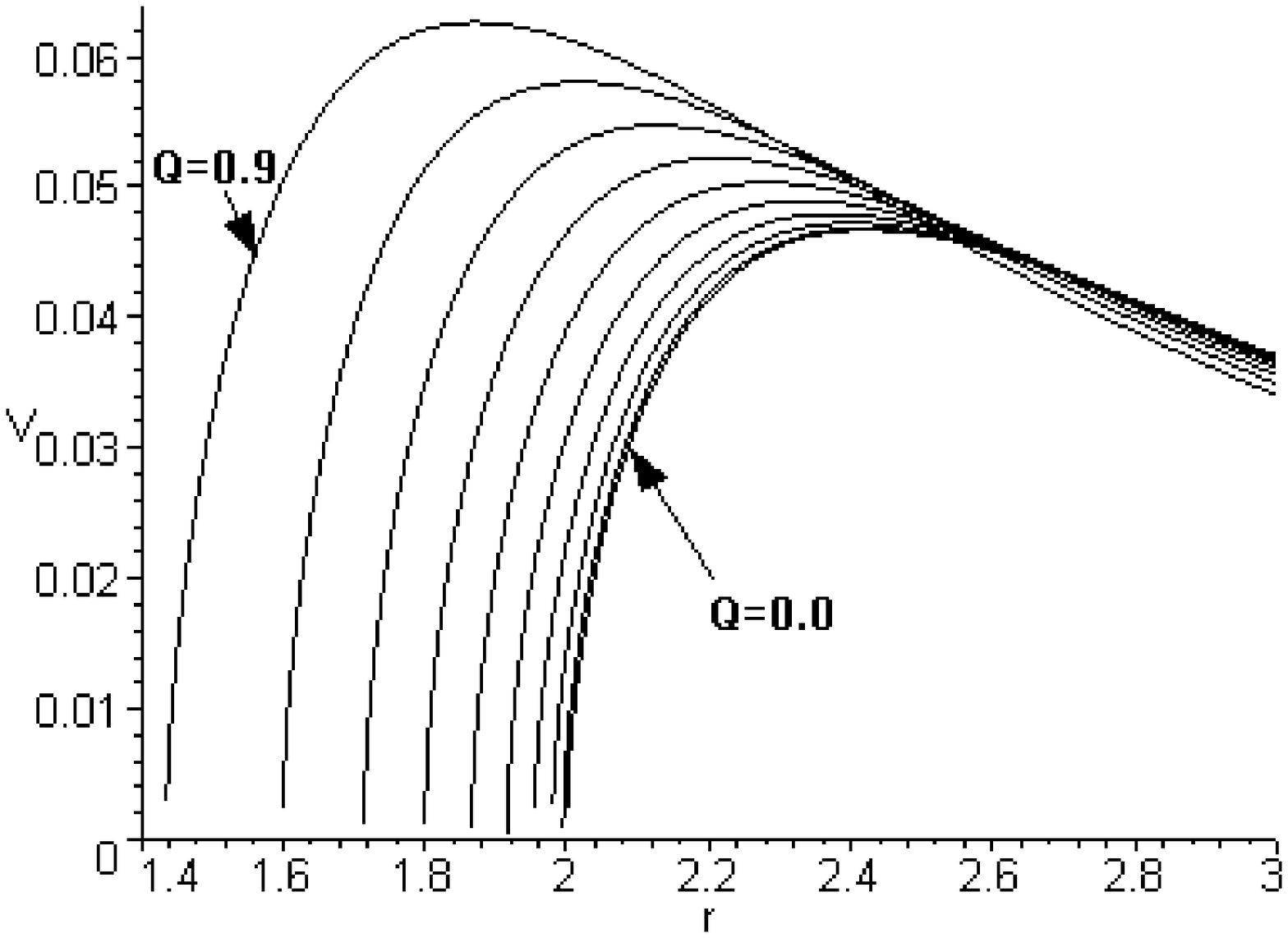}}
\caption{\label{Q1}Variation of the effective potential $V$ with
$Q$ which increases from right to left($0.0,0.1,\cdots,\ 0.9$) in
the case of $M=1,\lambda=1$.}
\end{center}
\end{figure}

\begin{figure}
\begin{center}
\scalebox{0.8}[0.8]{\includegraphics{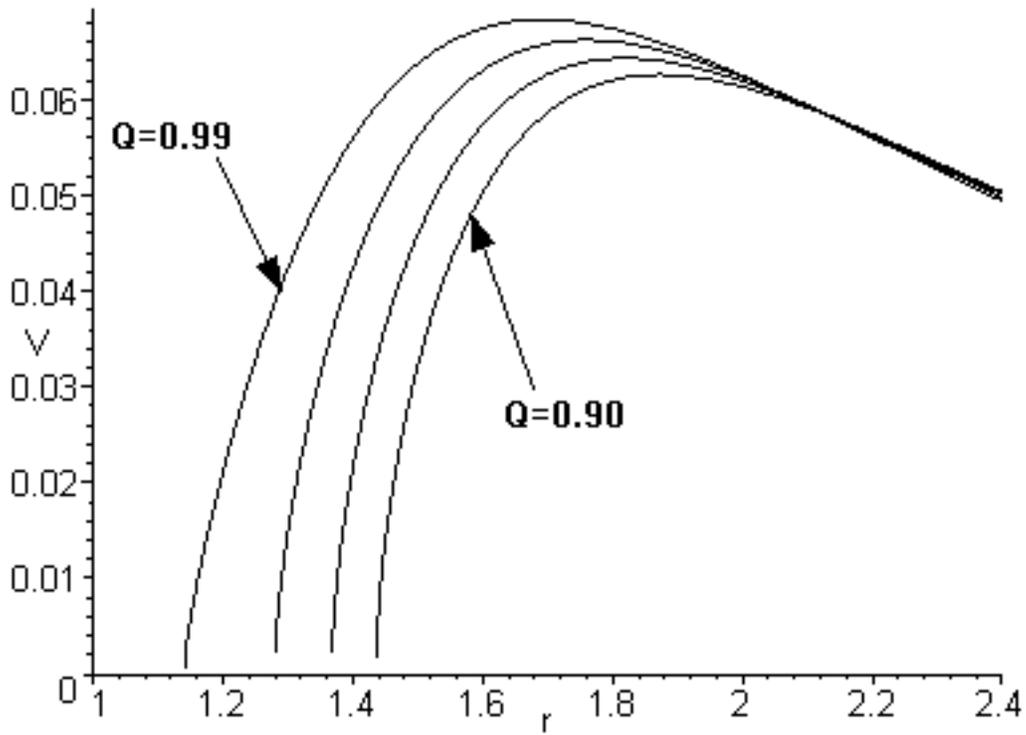}}
\caption{\label{Q2}Variation of the effective potential $V$ with
$Q$ which increases from right to left($0.90,0.93,0.96,0.99$) in
the case of $M=1,\lambda=1$.}
\end{center}
\end{figure}

As is shown in Fig.~\ref{Q1} and Fig.~\ref{Q2}, the above
effective potential $V\rightarrow 0$ with $r_*\rightarrow
-\infty$($r\rightarrow r_+$) or $r_*\rightarrow
\infty$($r\rightarrow \infty$); in addition, this effective
potential $V$ rises to its maximum at $r_*(r_{max})$. Hence the
QNM can be calculated by WKB approach, with the boundary condition
of purely "outgoing" waves. The formula for the spectra of QNM in
WKB approximation, carried to the third order beyond the eikonal
approximation, is given by\cite{IW}
\begin{equation}
\omega^{2}=[V_{0}+(-2V^{''}_{0})^{1/2}\Lambda]
-i(n+\frac{1}{2})[(1+\Omega)] \label{WKBeq}
\end{equation}
where
\begin{eqnarray}
\Lambda&=&\frac{1}{(-2V^{''}_{0})^{1/2}}
\left\{\frac{1}{8}\left(\frac{V_{0}^{(4)}}{V_{0}^{''}}\right)
\left(\frac{1}{4}+\alpha^{2}\right)- \frac{1}{288}
\left(\frac{V_{0}^{'''}}{V_{0}^{''}}\right)^{2}(7+60\alpha^{2})\right\}\\
\Omega&=&\frac{1}{(-2V_{0}^{''})}\left\{\frac{5}{6912}
\left(\frac{V_{0}^{'''}}{V_{0}^{''}}\right)^{4}(77+188\alpha^{2})
-\frac{1}{384}\left(\frac{{V_{0}^{'''}}^{2}V_{0}^{(4)}}{{V_{0}^{''}}^{3}}\right)
(51+100\alpha^{2})\right.\nonumber\\ &&\ \ \
+\frac{1}{2304}\left(\frac{V_{0}^{(4)}}{V_{0}^{''}}\right)^{2}(67+68\alpha^{2})
+\frac{1}{288}\left(\frac{V_{0}^{'''}V_{0}^{(5)}}{{V_{0}^{''}}^{2}}\right)(19+28\alpha^{2})
\nonumber\\ &&\ \ \ \left.
-\frac{1}{288}\left(\frac{V_{0}^{(6)}}{V_{0}^{''}}\right)(5+4\alpha^{2})\right\}
\end{eqnarray}
here
\begin{eqnarray}
\alpha&=&n+\frac{1}{2},\ n=\left\{
\begin{array}{l}
0,1,2,\cdots,\ {\rm Re}(\omega)>0\\ -1,-2,-3,\cdots,\ {\rm
Re}(\omega)<0
\end{array}
\right.\\
V_{0}^{(n)}&=&\left.\frac{d^{n}V}{dr_{\ast}^{n}}\right|_{r_{\ast}=r_{\ast}(r_{max})}
\end{eqnarray}
According to this analytic approximation formula, we find that the
real parts of the frequencies $\rm{Re}(\omega)$ are mainly
determined by the maximum of the effective potential $V$: the more
the maximum of the effective potential is, the more the real parts
of the frequencies are; the imaginary parts $\rm{Im}(\omega)$
mainly by the mode number $n$: with the mode number $n$ on the
increase, the imaginary parts decrease fastly. This last point
indicates that the QNM with higher mode number $n$ will decay
faster than the low-lying modes. Hence, the low-lying QNM are the
most relevant to the description of the evolution of fields around
black holes. This approximation formula has been used extensively
in various cases and is good enough for our purpose: comparing
with numerical results, this approximation has been found to be
accurate up to around 1\% for both the real and the imaginary
parts of the frequencies for low-lying modes with
$n<\lambda$\cite{Cho}.
\begin{table}
\begin{center}
\begin{tabular}{|c|c|c|c|}
  \hline
  M=1 & Q=0.0 & Q=0.3 & Q=0.6 \\
  \hline

 $\lambda=1$, $n=0$& $0.1765-0.1001i$& $0.1799-0.1005i$& $0.1919-0.1012i$\\
 $\lambda=2$, $n=0$& $0.3786-0.0965i$& $0.3847-0.0970i$& $0.4059-0.0981i$\\
 $\lambda=2$, $n=1$& $0.3536-0.2987i$& $0.3602-0.2999i$& $0.3834-0.3023i$\\
 $\lambda=3$, $n=0$& $0.5737-0.0963i$& $0.5827-0.0968i$& $0.6140-0.0979i$\\
 $\lambda=3$, $n=1$& $0.5562-0.2930i$& $0.5655-0.2943i$& $0.5982-0.2973i$\\
 $\lambda=3$, $n=2$& $0.5273-0.4972i$& $0.5372-0.4992i$& $0.5719-0.5037i$\\
 $\lambda=4$, $n=0$& $0.7672-0.0963i$& $0.7792-0.0967i$& $0.8208-0.0979i$\\
 $\lambda=4$, $n=1$& $0.7540-0.2910i$& $0.7662-0.2924i$& $0.8088-0.2956i$\\
 $\lambda=4$, $n=2$& $0.7304-0.4909i$& $0.7431-0.4930i$& $0.7874-0.4980i$\\
 $\lambda=4$, $n=3$& $0.6999-0.6957i$& $0.7131-0.6986i$& $0.7595-0.7051i$\\
 $\lambda=5$, $n=0$& $0.9602-0.0963i$& $0.9752-0.0967i$& $1.0272-0.0979i$\\
 $\lambda=5$, $n=1$& $0.9496-0.2902i$& $0.9647-0.2916i$& $1.0175-0.2949i$\\
 $\lambda=5$, $n=2$& $0.9300-0.4876i$& $0.9455-0.4899i$& $0.9996-0.4951i$\\
 $\lambda=5$, $n=3$& $0.9036-0.6892i$& $0.9196-0.6923i$& $0.9755-0.6992i$\\
 $\lambda=5$, $n=4$& $0.8721-0.8944i$& $0.8886-0.8982i$& $0.9468-0.9066i$\\
 \hline

\end{tabular}
\begin{tabular}{|c|c|c|c|}
  \hline
  M=1 &  Q=0.7 & Q=0.8 & Q=0.9 \\
  \hline

 $\lambda=1$, $n=0$& $0.1990-0.1013i$& $0.2083-0.1006i$& $0.2207-0.0979i$\\
 $\lambda=2$, $n=0$& $0.4182-0.0984i$& $0.4348-0.0981i$& $0.4581-0.0963i$\\
 $\lambda=2$, $n=1$& $0.3969-0.3025i$& $0.4150-0.3009i$& $0.4396-0.2944i$\\
 $\lambda=3$, $n=0$& $0.6322-0.0982i$& $0.6568-0.0980i$& $0.6915-0.0963i$\\
 $\lambda=3$, $n=1$& $0.6172-0.2979i$& $0.6428-0.2969i$& $0.6787-0.2913i$\\
 $\lambda=3$, $n=2$& $0.5922-0.5042i$& $0.6194-0.5019i$& $0.6566-0.4914i$\\
 $\lambda=4$, $n=0$& $0.8450-0.0981i$& $0.8776-0.0980i$& $0.9239-0.0963i$\\
 $\lambda=4$, $n=1$& $0.8336-0.2963i$& $0.8670-0.2955i$& $0.9142-0.2902i$\\
 $\lambda=4$, $n=2$& $0.8132-0.4988i$& $0.8480-0.4971i$& $0.8966-0.4875i$\\
 $\lambda=4$, $n=3$& $0.7866-0.7059i$& $0.8232-0.7029i$& $0.8731-0.6884i$\\
 $\lambda=5$, $n=0$& $1.0573-0.0981i$& $1.0981-0.0979i$& $1.1559-0.0963i$\\
 $\lambda=5$, $n=1$& $1.0481-0.2956i$& $1.0896-0.2949i$& $1.1481-0.2897i$\\
 $\lambda=5$, $n=2$& $1.0311-0.4961i$& $1.0738-0.4946i$& $1.1336-0.4854i$\\
 $\lambda=5$, $n=3$& $1.0082-0.7003i$& $1.0523-0.6977i$& $1.1135-0.6841i$\\
 $\lambda=5$, $n=4$& $0.9808-0.9078i$& $1.0266-0.9040i$& $1.0892-0.8855i$\\
 \hline

\end{tabular}
\end{center}
\caption{ QNM frequencies for Weyl neutrino field in R-N black
holes.\label{table1}}
\end{table}
\begin{figure}
\begin{center}
\scalebox{1.6}[1.6]{\includegraphics{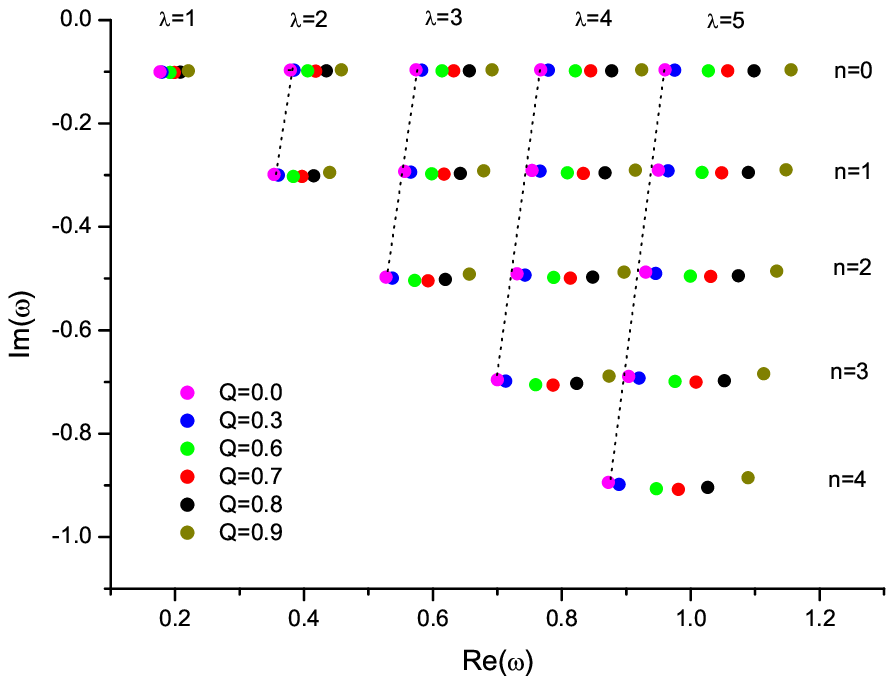}}
\caption{\label{fig3}QNM frequencies for Weyl neutrino field in
R-N black holes.}
\end{center}
\end{figure}
\begin{table}
\begin{center}
\begin{tabular}{|c|c|c|c|c|}
\hline
  M=1 & Q=0.90 & Q=0.93 & Q=0.96 & Q=0.99 \\
  \hline

 $\lambda=1$, $n=0$& $0.2207-0.0979i$& $0.2250-0.0962i$& $0.2294-0.0936i$& $0.2336-0.0902i$\\
 $\lambda=2$, $n=0$& $0.4581-0.0963i$& $0.4670-0.0950i$& $0.4773-0.0930i$& $0.4891-0.0898i$\\
 $\lambda=2$, $n=1$& $0.4396-0.2944i$& $0.4485-0.2901i$& $0.4579-0.2836i$& $0.4668-0.2740i$\\
 $\lambda=3$, $n=0$& $0.6915-0.0963i$& $0.7050-0.0951i$& $0.7206-0.0931i$& $0.7390-0.0899i$\\
 $\lambda=3$, $n=1$& $0.6787-0.2913i$& $0.6923-0.2873i$& $0.7076-0.2812i$& $0.7242-0.2715i$\\
 $\lambda=3$, $n=2$& $0.6566-0.4914i$& $0.6701-0.4844i$& $0.6841-0.4737i$& $0.6971-0.4578i$\\
 $\lambda=4$, $n=0$& $0.9239-0.0963i$& $0.9419-0.0951i$& $0.9628-0.0931i$& $0.9876-0.0899i$\\
 $\lambda=4$, $n=1$& $0.9142-0.2902i$& $0.9324-0.2864i$& $0.9531-0.2805i$& $0.9767-0.2708i$\\
 $\lambda=4$, $n=2$& $0.8966-0.4875i$& $0.9148-0.4808i$& $0.9349-0.4705i$& $0.9558-0.4543i$\\
 $\lambda=4$, $n=3$& $0.8731-0.6884i$& $0.8910-0.6786i$& $0.9097-0.6638i$& $0.9265-0.6414i$\\
 $\lambda=5$, $n=0$& $1.1559-0.0963i$& $1.1784-0.0951i$& $1.2046-0.0932i$& $1.2358-0.0899i$\\
 $\lambda=5$, $n=1$& $1.1481-0.2897i$& $1.1708-0.2860i$& $1.1969-0.2802i$& $1.2271-0.2705i$\\
 $\lambda=5$, $n=2$& $1.1336-0.4854i$& $1.1564-0.4790i$& $1.1821-0.4689i$& $1.2102-0.4527i$\\
 $\lambda=5$, $n=3$& $1.1135-0.6841i$& $1.1363-0.6746i$& $1.1611-0.6601i$& $1.1860-0.6375i$\\
 $\lambda=5$, $n=4$& $1.0892-0.8855i$& $1.1117-0.8729i$& $1.1349-0.8538i$& $1.1555-0.8251i$\\
 \hline
\end{tabular}
\end{center}
\caption{ QNM frequencies for Weyl neutrino field in near extremal
R-N black holes.\label{table2}}
\end{table}
\begin{figure}
\begin{center}
\scalebox{1.6}[1.6]{\includegraphics{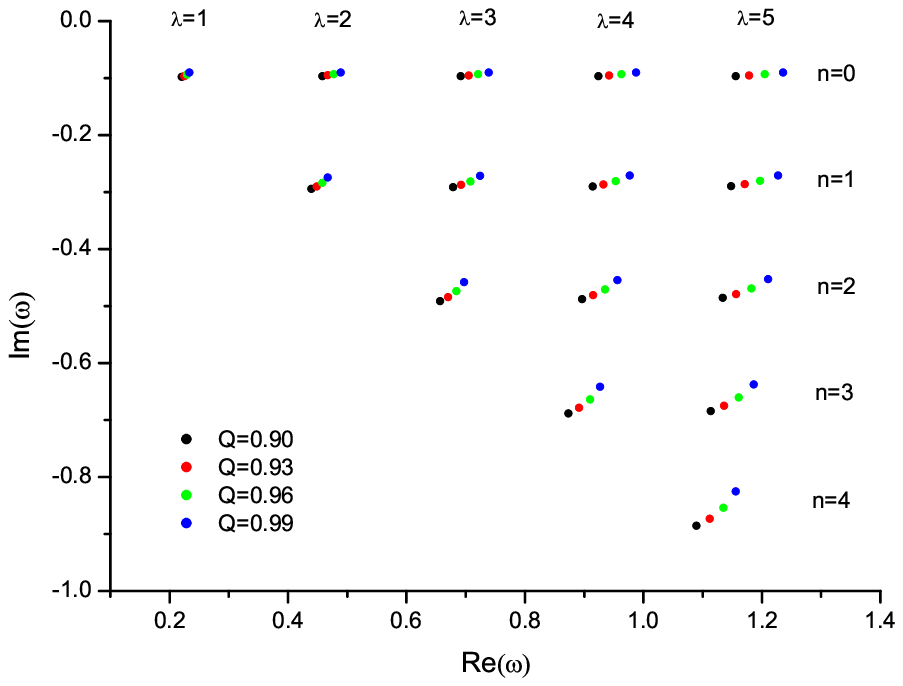}}
\caption{\label{fig4}QNM frequencies for Weyl neutrino field in
near extremal R-N black holes.}
\end{center}
\end{figure}

To proceed, it is convenient to take the mass of the R-N black
hole considered as a unit of mass in geometrized units. Later
plugging the effective potential $V$ in (\ref{V}) into the
approximation formula above, we obtain the complex QNM frequencies
for Weyl neutrino field in R-N black holes. The values for $0\leq
n<\lambda\leq 5$ are listed in Table ~\ref{table1}, Table
~\ref{table2}, and plotted in Fig.~\ref{fig3}, Fig.~\ref{fig4}
correspondingly.

Hereby we find the main results as follow.

 I. QNM for Weyl
neutrino field in Charged black holes show their deviations from
Schwarzchild black holes. For $\lambda$ and $n$ fixed, with the
charge $Q$ on the increase, the real parts $\rm{Re}(\omega)$ of
the frequencies increase monotonously, which agrees with
Fig.~\ref{Q1} and Fig.~\ref{Q2}: When the charge $Q$ increases,
the maximum of the effective potential $V$ also increases. In
comparison with the variation of the real parts of the
frequencies, their imaginary parts $\rm{Im}(\omega)$ seem to
change less with $Q$: decrease firstly, later fall to a minimum at
the vicinity of $Q=0.7$, and increase finally. This result shows
that the charge-induced effective gravitational potential has an
influence on the QNM for Weyl neutrino field, although neutrino
does not take part in the charge-charge electromagnetic
interaction.

 II. In addition, as for
a definite $Q$, the real parts $\rm{Re}(\omega)$ of the
frequencies increase with $\lambda$ for a fixed $n$\footnote{This
result is physically reasonable, because the larger $\lambda$ is,
the larger the corresponding effective centrifugal potential is.
}, and the $\rm{Im}(\omega)$ almost remain constant; on the other
hand, the $\rm{Re}(\omega)$ decrease slowly as the mode number $n$
increases for the same $\lambda$, at the same time, the imaginary
parts $\rm{Im}(\omega)$ decrease with $n$ fastly, which means the
low-lying QNM dominate in the intermediate evolution stage of the
field.

III. Despite its spin half, the above variations of QNM
frequencies for Weyl neutrino field are the same as those for
integral spin fields such as scalar, electromagnetic and
gravitational fields in R-N black holes\cite{KS1,Konoplya1}.
\section{Conclusion and discussion}
We have calculated the low-lying QNM frequencies for Weyl neutrino
field in R-N black holes using WKB approximation up to the third
order. The result shows that the charge-induced effective
gravitation makes QNM for Weyl neutrino field in Charged black
holes different from Schwarzchild black holes, and the variations
of QNM frequencies for Weyl neutrino field are similar to those
for integral spin fields in R-N black holes.

We conclude with two interesting problems worthy of further
investigation: One is the calculation of the highly damped
asymptotic QNM for Weyl neutrino field in R-N black holes, which
is outside our consideration but the most relevant to the quantum
charged black holes.  The other is the search of low-lying QNM
behaviors for charged fermion field in R-N black holes as charged
scalar field\cite{Konoplya1}, where the QNM will be influenced by
not only the charge-induced gravitation but also the charge-charge
electromagnetic force. One expected result is that the real parts
of the QNM for the fermion field with the same charge as that of a
R-N black hole will be larger than the neutral fermion field, due
to the repulsive electromagnetic interaction, which leads to the
rise of the maximum of the effective potential.
\section*{Acknowledgement}
It is our pleasure to acknowledge Prof. S. Pei for the discussion
of neutrino astrophysics. H. Zhang would like to thank Prof. H. T.
Cho for many helpful arguments on this work. In addition, this
work is supported in part by NSFC(grant 10205002).

\end{document}